\documentclass[11pt]{article}
\usepackage{amsmath,amssymb}
\newcommand{\bra}[1]{\langle #1\vert}
\newcommand{\ket}[1]{\vert #1\rangle}
\newcommand{\braket}[2]{\langle #1\vert #2\rangle}

\newcommand{\iden}{{\bf 1}}
\newcommand{\Tr}{\mbox{Tr}}
\begin{document}

\title{Probability, Arrow of Time and Decoherence\thanks{To appear in {\em Studies in History and 
Philosophy of Modern Physics}.}}
\author{Guido Bacciagaluppi\footnote{Centre for Time, Department of Philosophy, University of Sydney,
Sydney NSW 2006, Australia (e-mail: guido.bacciagaluppi@arts.usyd.edu.au).}} 
\date{}
\maketitle

\begin{abstract}
This paper relates both to the metaphysics of probability and to the physics 
of time asymmetry. Using the formalism of decoherent histories, it investigates 
whether intuitions about intrinsic time directedness that are often associated with probability 
can be justified in the context of no-collapse approaches to quantum mechanics. 
The standard (two-vector) approach to time symmetry in the decoherent histories literature is criticised, and an alternative approach is proposed, based on two decoherence conditions (`forwards' and `backwards') within the one-vector formalism. In turn, considerations of forwards and backwards decoherence and of decoherence and recoherence suggest that a time-directed interpretation of probabilities, if adopted, should be both contingent and perspectival.
\end{abstract}

\section{Introduction}\label{probs}

Probabilities are often treated in a time-asymmetric way, for example when used in deliberations about
future courses of action, or when chances are said to evolve by conditionalisation upon events being 
actualised, so that chances of past events are always equal to 0 or 1. In metaphysical terms, this is
often associated with the notion of an `open future' and a `fixed past'. Is this asymmetry intrinsic to the notion of probability, or is it imposed at the level of interpretation? This is the main question we wish to examine in this paper, although only in a special case.

The general strategy we suggest to adopt in order to address this question is to investigate whether some kind of time asymmetry is already present at the level of the formalism. A formal investigation of course can yield no normative conclusion about the interpretation of the quantum probabilities. However, we take it that it can provide useful guidelines for choosing or constructing a good interpretation. For instance, one could consider probabilities as used in the description of classical stochastic processes. We claim (but will not discuss this point here) that in this example there is no formal equivalent of an open future, since a classical stochastic process is just a probability measure over a space of trajectories, so the formal definition is completely time-symmetric. Transition probabilities
towards the future can be obtained by conditionalising on the past; equally, transition probabilities
towards the past can be obtained by conditionalising on the future. Individual trajectories may 
exhibit time asymmetry, and there may be a quantitative asymmetry between forwards and backwards transition probabilities, but at least as long as the latter are not all 0 or 1, this falls well short of justifying a 
notion of fixed past.

In this paper, we shall apply this general strategy in a different context, namely that of no-collapse approaches to quantum mechanics. For vividness's sake, one can imagine that we are discussing an Everett-like interpretation, for instance one in which `worlds' are identified with the `histories' of the decoherent histories approach (briefly sketched below), and in which the probabilities emerge from the deterministic Schr\"{o}dinger equation as objective chances identified through a decision-theoretic analysis (see for instance Saunders 1993 and Wallace 2005 for these two aspects, respectively). However, and I wish to emphasise this point from the outset, our analysis will be carried out purely at the formal level, so that our results may be applied to the quantum probabilities irrespective of the chosen (no-collapse) approach or of the attendant interpretation of probability. In particular, I believe that the analysis below will apply also to pilot-wave theories, such as (deterministic) de Broglie-Bohm theory, where probabilities emerge in a way roughly analogous to that in classical statistical mechanics (see e.g.\ D\"{u}rr, Goldstein and Zangh\`{\i} 1992), or such as (stochastic) beable theories, where the notion of probability is presupposed and its interpretation, presumably, is open (Bell 1984).  

The way we shall proceed is as follows. Probabilities enter quantum mechanics at the level of the (real or apparent) collapse of the quantum state, the description given by the Schr\"{o}dinger evolution being entirely deterministic. Time asymmetry also enters quantum mechanics at the level of collapse, the Schr\"{o}dinger equation being time-symmetric in a well-defined sense. The quantum probabilities thus appear to be very good candidates for probabilities that are intrinsically time-directed. We shall sketch the relevant aspects of these questions in section~\ref{quantum}.

Then we shall focus on no-collapse approaches to quantum mechanics. In order to keep the discussion general, and not get tied up with questions of interpretation, we shall discuss probabilities and their putative time-directedness within the framework provided by decoherence, specifically by the formalism of decoherent histories. I have suggested elsewhere (Bacciagaluppi 2003) that the phenomenon of decoherence is the crucial ingredient that allows all major no-collapse interpretations to recover the appearance of collapse. The argument below does not depend on this, although of course 
its range of application does: the argument will apply precisely to those approaches to quantum mechanics (and their associated notion of probability) that make such use of decoherence. The idea of decoherence is often linked to the picture of a branching structure for the universal wave function, 
which in turn (at least from an Everettian perspective) is close to the intuition of `open future', but  
our discussion has no need to link the histories formalism and its probabilities to either
Everett interpretations or any other interpretations of quantum mechanics.

The notion of decoherence will be briefly sketched in section~\ref{histories}, together with the specific formalism of decoherent histories. Our use of decoherent histories does not carry any particular interpretational commitments, despite the amount of controversy that has surrounded this notion. 
We have chosen this formalism simply because it provides a very convenient
and explicit way of discussing probabilities in the framework of decoherence. In particular, decoherent histories allow us to embed collapse-style probabilities, with their time-asymmetric aspects, within the no-collapse framework of the Schr\"{o}dinger equation. 

In section~\ref{asymmetry}, we shall criticise a common misconception regarding the arrow of time in 
the decoherent histories formalism. Instead of introducing a two-vector formalism, as is common in the literature, we shall retain the usual one-vector formalism of the Schr\"{o}dinger equation and introduce separate `forwards' and `backwards' decoherence conditions. Our proposal was 
sketched already in Bacciagaluppi (2002) (which focuses on the related `branching 
space-time' structure of the wave function and uses it to discuss the concept of locality in
many-worlds interpretations). 

We shall use our conclusions to discuss the status of probabilities in no-collapse quantum 
mechanics. This will take up our final section~\ref{discussion}. 
On the basis of our discussion of forwards and backwards decoherence from section~\ref{asymmetry}, we shall suggest that an interpretation of the no-collapse quantum probabilities should allow for time directedness to be a merely contingent feature of the probabilities. On the other hand, we shall argue that decoherence can only be observed as decoherence and never as recoherence from the perspective of an internal observer, so that there is scope for time directedness in the interpretation of probabilities, but in a perspectival sense.

\section{Quantum mechanics and time (a)symmetry}\label{quantum}

\subsection{Schr\"{o}dinger equation and arrow of collapse}\label{arrow}

The general form of the problem of the arrow of time in physics is that the fundamental equations are assumed to be time-symmetric, but that some class of phenomena appear to be time-directed (or at least disproportionately favouring one time direction). An example specific to quantum mechanics is the tension between the time symmetry of the Schr\"{o}dinger equation and the time asymmetry of what phenomenologically appears as `collapse'.

Time reversal for wave
functions is implemented as
  \begin{equation}
    \psi(x,t)\mapsto\psi^*(x,-t)\ ,
    \label{1}
  \end{equation}
and thus, because $H^*=H$, the Schr\"{o}dinger equation
  \begin{equation}
    i\hbar\frac{\partial}{\partial t}\psi(x,t)=H\psi(x,t)
    \label{2}
  \end{equation}
is time-symmetric.

It is quite obvious that (\ref{2}) allows for solutions that are not individually time-symmetric,
and thus allows for time-asymmetric behaviour. Indeed, a time-symmetric solution 
$\psi(x,t)$ will be symmetric iff
  \begin{equation}
     \psi(x,t)=\psi^*(x,-t)\ .
     \label{4}
  \end{equation}
This is a non-trivial condition, equivalent to having
  \begin{equation}
     \psi(x,0)=\psi^*(x,0)
     \label{5}
  \end{equation}
for some time $t=0$. 

However, the problem of the `arrow of collapse' is that some phenomena, specifically in situations of measurement, appear not to be described at all by the Schr\"{o}dinger equation, but by the so-called collapse or projection postulate, which in the simplest case states the following. At measurement times, a state $\ket{\psi}$ appears to be transformed to one of the states $P_\alpha\ket{\psi}$ (up to normalisation), with probability $\mbox{Tr}(\ket{\psi}\bra{\psi}P_\alpha)$, where the $P_\alpha$ are the 
eigenprojections of the measured observable. Therefore the (apparent) evolution of a quantum state $\ket{\psi(t_0)}$ is given alternatively by periods of Schr\"{o}dinger evolution $U_{t_{i+1}t_i}$ and 
`collapses' $P_{\alpha_i}$:
  \begin{equation}
    \ket{\psi(t_0)} \longrightarrow U_{t_{n+1}t_n}P_{\alpha_n}U_{t_nt_{n-1}}\ldots P_{\alpha_1}U_{t_1t_0}\ket{\psi (t_0)} 
    \label{new6}
  \end{equation}
(up to normalisation), with overall probability 
  \begin{equation}
    \Tr\Big(P_{\alpha_n}U_{t_nt_{n-1}}\ldots P_{\alpha_1}U_{t_1t_0}\ket{\psi (t_0)}     
            \bra{\psi (t_0)}U^*_{t_1t_0}P_{\alpha_1}\ldots U^*_{t_nt_{n-1}}P_{\alpha_n}\Big)\ .  
    \label{new7}
  \end{equation}
If we attempt a time reversal, taking the final state and applying to it the same procedure 
in reverse,
  \begin{equation}
    \ket{\psi(t_{n+1})} \longrightarrow U^*_{t_1t_0}P_{\alpha_1}U^*_{t_2t_1}\ldots P_{\alpha_n}U^*_{t_{n+1}t_n}\ket{\psi (t_{n+1})} 
    \label{new8}
  \end{equation}
(again up to normalisation), we see that in general this state is different from $\ket{\psi (t_0)}$ (nor is the
sequence of intermediate states reversed). The corresponding probability, 
  \begin{equation}
    \Tr\Big(P_{\alpha_1}U^*_{t_2t_1}\ldots P_{\alpha_n}U^*_{t_{n+1}t_n}\ket{\psi (t_{n+1})}     
            \bra{\psi (t_{n+1})}U_{t_{n+1}t_n}P_{\alpha_n}\ldots U_{t_2t_1}P_{\alpha_1}\Big)\ ,  
    \label{new9}
  \end{equation}
in general is also different from (\ref{new7}). 

As a simple example take a spin-1/2 particle in an arbitrary initial state 
$\ket{\psi(t_0)}=\alpha\ket{+_x}+\beta\ket{-_x}$, and let it be subject to a measurement of spin in $x$-direction 
at time $t_1$, and to a measurement of spin in $z$-direction at time $t_2$, with no further evolution afterwards.

The sequence of states from $t_0$ to $t_3$ is then for instance given (up to normalisation) by:
  \begin{multline}
      (\alpha\ket{+_x}+\beta\ket{-_x})\otimes\ket{M^1_0}\otimes\ket{M^2_0}   \\
    \begin{split}
        & \xrightarrow{U_{t_1t_0}} 
        & \alpha\ket{+_x}\otimes\ket{M^1_+}\otimes\ket{M^2_0}+\beta\ket{-_x}\otimes\ket{M^1_-}\otimes\ket{M^2_0} 
      \\
        & \xrightarrow{P^x_+}
        & \alpha\ket{+_x}\otimes\ket{M^1_+}\otimes\ket{M^2_0}     
      \\
        & \xrightarrow{U_{t_2t_1}}
        & \frac{\alpha}{\sqrt{2}}\ket{+_z}\otimes\ket{M^1_+}\otimes\ket{M^2_+}+
          \frac{\alpha}{\sqrt{2}}\ket{-_z}\otimes\ket{M^1_+}\otimes\ket{M^2_-}
      \\
        & \xrightarrow{P^z_+}
        & \frac{\alpha}{\sqrt{2}}\ket{+_z}\otimes\ket{M^1_+}\otimes\ket{M^2_+}
      \\
        & \xrightarrow{U_{t_3t_2}}
        & \frac{\alpha}{\sqrt{2}}\ket{+_z}\otimes\ket{M^1_+}\otimes\ket{M^2_+}\ ,
    \end{split}
    \label{Paris1}
  \end{multline}
and the corresponding probability for the event `spin-$x$ up, followed by spin-$z$ up' is $\frac{|\alpha|^2}{2}$.

The reverse procedure, however, yields the following (always up to normalisation):
  \begin{equation}
    \begin{array}{clr}
      \ket{+_z}\otimes\ket{M^1_+}\otimes\ket{M^2_+}  
        & \xrightarrow{U^*_{t_3t_2}}
        & \ket{+_z}\otimes\ket{M^1_+}\otimes\ket{M^2_+}
      \\
        & \xrightarrow{P^{z*}_+=P^z_+}
        & \ket{+_z}\otimes\ket{M^1_+}\otimes\ket{M^2_+}
      \\
        & \xrightarrow{U^*_{t_2t_1}}
        & \ket{+_z}\otimes\ket{M^1_+}\otimes\ket{M^2_0}
      \\
        & \xrightarrow{P^{x*}_+=P^x_+}
        & \frac{1}{\sqrt{2}}\ket{+_x}\otimes\ket{M^1_+}\otimes\ket{M^2_0}
      \\
        & \xrightarrow{U^*_{t_1t_0}}
        & \frac{1}{\sqrt{2}}\ket{+_x}\otimes\ket{M^1_+}\otimes\ket{M^2_0}\ ,
\end{array}
    \label{Paris2}
  \end{equation}
with a probability for `spin-$z$ up, preceded by spin-$x$ up' of $\frac{1}{2}$.

Therefore, the apparent time evolution of the state, i.e.\ the evolution including collapse, is clearly not 
time-symmetric, as opposed to that given by the Schr\"{o}dinger evolution.

\subsection{Time symmetry in collapse approaches and no-collapse approaches to quantum mechanics}\label{collnocoll}
The main alternative in the foundational approaches to quantum mechanics is the alternative 
between collapse approaches, which take (\ref{new6}), or rather, some more precise variant thereof, as fundamental, and 
no-collapse approaches, which take the Schr\"{o}dinger equation as fundamental, and attempt to explain  (\ref{new6}) as 
some effective description. 

If one wishes to adopt the first alternative and at the same time 
uphold time symmetry at the level of the fundamental equations, then one is committed to a symmetrisation of the 
collapse postulate. To this end, one can use the well-known symmetrisation of 
the probability formula (\ref{new7}) introduced by Aharonov, Bergmann and Lebowitz (1964), where instead of the 
probability formula (\ref{new7}) one uses the so-called ABL formula:
  \begin{equation}
    \frac{\begin{split}\Tr\Big(\ket{\psi (t_{n+1})}\bra{\psi (t_{n+1})}
                  U_{t_{n+1}t_n}P_{\alpha_n}U_{t_nt_{n-1}}\ldots P_{\alpha_1}U_{t_1t_0}\\[-1ex] 
                  \ket{\psi (t_0)}\bra{\psi (t_0)}
                  U^*_{t_1t_0}P_{\alpha_1}\ldots U^*_{t_nt_{n-1}}P_{\alpha_n}U^*_{t_{n+1}t_n}\Big)\end{split}
    }{\begin{split}\sum_{\alpha_1,\ldots,\alpha_n}
          \Tr\Big(\ket{\psi (t_{n+1})}\bra{\psi (t_{n+1})}
                  U_{t_{n+1}t_n}P_{\alpha_n}U_{t_nt_{n-1}}\ldots P_{\alpha_1}U_{t_1t_0}\\[-2ex] 
                  \ket{\psi (t_0)}\bra{\psi (t_0)}
                  U^*_{t_1t_0}P_{\alpha_1}\ldots U^*_{t_nt_{n-1}}P_{\alpha_n}U^*_{t_{n+1}t_n}\Big)\end{split}
    }\ .
    \label{new10}
  \end{equation}

Note that if one takes as $\ket{\psi (t_{n+1})}$ one of the possible `collapsed' states after a sequence of
measurements, this formula can be understood simply as describing post-selection in an ensemble of systems. This use
of the formula is compatible with any approach to quantum mechanics, even with a no-collapse Everett approach (as in 
Vaidman's work, see e.g.\ Vaidman 2007), or indeed with the usual time-asymmetric, forward-in-time collapse postulate.

On the other hand, one can use this formula to define an explicitly time-symmetric collapse approach, by taking 
$\ket{\psi (t_0)}$ and $\bra{\psi (t_{n+1})}$ to be independent of each other. These two quantum states can both be understood 
as evolving according to the unitary evolution interrupted by occasional collapses, one towards the future and
one towards the past. At each instant there are therefore two quantum states, and both contribute, via (\ref{new10}), 
to the probability for the collapses that punctuate their evolutions. A two-vector formalism based on the ABL formula 
(\ref{new10}) with arbitary $\ket{\psi (t_0)}$ and $\bra{\psi (t_{n+1})}$ has been explicitly proposed by Aharonov and
Vaidman (2002). A symmetrised collapse is presumably the natural picture behind this formalism.

In this paper we focus on the second alternative. In a no-collapse approach to the arrow of collapse, one will take the 
time-symmetric Schr\"{o}dinger equation, 
as described above, to be the fundamental equation of the theory, and seek to explain time asymmetry purely as a feature 
of effective collapse. The appearance of collapse will be explained differently in the different approaches
(by adding Bohm corpuscles for instance, or by interpreting the quantum state as describing many worlds or minds), and 
the topic of this paper are those no-collapse approaches that see the appearance of collapse as 
somehow related to loss of interference as described by the theory of decoherence. We thus turn to a brief description
of the latter.

\section{Decoherence and decoherent histories}\label{histories}
Loosely speaking, decoherence is the suppression of interference between components of the state of a system (or of some 
degrees of freedom of a system) through suitable interaction with the environment (or with some other degrees of freedom
of the system). A paradigm example is a two-slit experiment with sufficiently strong light shining on the electron 
between the slits and the screen: 
if photons are scattered off the electrons (thus, in a sense, detecting the passage of the electrons through either slit), 
the interference pattern at the screen is suppressed. If this is the case, 
we can also talk about which `trajectory' an individual electron has followed and we can consistently assign 
probabilities to alternative trajectories, so that probabilities for detection at the screen can be calculated 
by summing over intermediate events. None of this strictly formal talk of probabilities of course implies that the
electron has actually gone through one or other of the slits (or, when decoherence is applied to measurement situations, 
that the measurement yields a definite result).

The decoherent histories formalism (Griffiths 1984, Omn\`{e}s 1988, Gell-Mann and Hartle 1990), provides an abstract 
approach to decoherence 
by indeed defining it in terms of when we can obtain consistent formal expressions for the probabilities of alternative histories, defined 
in turn as time-ordered sequences of projection operators (usually Heisenberg-picture operators), strictly within the formal apparatus of no-collapse quantum mechanics. While the various 
attempts at basing interpretations of quantum mechanics around this formalism are the subject of  
controversy (see Dowker and Kent 1996), we take it that the {\em formalism} of decoherent histories can be used uncontroversially 
as an abstract description of certain features of decoherence. These features include the possibility of defining over time the 
identity of components of the state and of formally defining collapse-type probabilities. This is the aspect that is of 
particular interest here.

The formalism can be described in a nutshell as follows. Take orthogonal families of projections with
  \begin{equation}
    \sum_{\alpha_1}P_{\alpha_1}(t)={\bf 1},\ldots,
    \sum_{\alpha_n}P_{\alpha_n}(t)={\bf 1}
    \label{6}
  \end{equation}
(in Heisenberg picture). Choose times $t_1,\ldots,t_n$ and define {\em histories} as time-ordered sequences of projections
  \begin{equation}
    P_{\alpha_1}(t_1),\ldots,P_{\alpha_n}(t_n)
    \label{7}
  \end{equation}
at the given times, choosing one projection from each family, respectively. The histories are then said to form an alternative 
and exhaustive set.

Given a state $\rho$, we wish to define probabilities for the resulting set (including further coarse-grainings of the 
histories). The usual probability formula based on the collapse postulate would yield
  \begin{equation}
    \mbox{Tr}\Big( P_{\alpha_n}(t_n)\ldots 
    P_{\alpha_1}(t_1)\rho P_{\alpha_1}(t_1)\ldots
    P_{\alpha_n}(t_n)\Big)
    \label{8}
  \end{equation}
as `candidate probabilities', 
but in general if one coarse-grains further, in the sense of summing over intermediate events, one does not obtain 
probabilities of the same form (\ref{8}). This is because in general there are non-zero interference terms between 
different histories. Now, the  {\em consistency} or (weak) {\em decoherence}
condition is precisely that interference terms should vanish
for any pair of distinct histories:
  \begin{equation}
    \mbox{Re}\mbox{Tr}\Big( P_{\alpha_n}(t_n)\ldots 
    P_{\alpha_1}(t_1)\rho P_{\alpha'_1}(t_1)\ldots
    P_{\alpha'_n}(t_n)\Big)=0\ 
    \label{9}
  \end{equation}
for $\{\alpha_i\}\neq\{\alpha'_i\}$ (we gloss over differences in the definitions adopted by various authors). In this case (\ref{8}) defines probabilities 
for all histories of the set (including coarse-grainings) or, equivalently, it defines the distribution functions for a 
stochastic process with the histories as trajectories.

Formulas (\ref{8}) and (\ref{9}) can easily be translated into the Schr\"{o}dinger picture, in which they read
  \begin{equation}
    \mbox{Tr}\Big( P_{\alpha_n}U_{t_nt_{n-1}}\ldots 
    P_{\alpha_1}U_{t_1t_0}\rho(t_0)U^*_{t_1t_0}         
    P_{\alpha_1}\ldots U^*_{t_nt_{n-1}}P_{\alpha_n}\Big)\ , 
    \label{8A}
  \end{equation}
and
  \begin{equation}
    \mbox{Re}\mbox{Tr}\Big( P_{\alpha_n}U_{t_nt_{n-1}}\ldots 
    P_{\alpha_1}U_{t_1t_0}\rho(t_0)                 
    U^*_{t_1t_0} P_{\alpha'_1}\ldots U^*_{t_nt_{n-1}}
    P_{\alpha'_n}\Big)=0\ ,
    \label{9A}
  \end{equation}
respectively.

One should also note that a stronger form of the decoherence condition, namely the vanishing of both the real and 
imaginary part of the trace expression in (\ref{9}), can 
be used to prove theorems on the existence of (later) `permanent records' of (earlier) events in a history. Indeed, if the
state $\rho$ is a pure state $\ket{\psi}\bra{\psi}$ this strong decoherence condition (sometimes also called `medium
decoherence') is equivalent, for all $n$, to the orthogonality of the vectors
  \begin{equation}
    P_{\alpha_n}(t_n)\ldots P_{\alpha_1}(t_1)\ket{\psi}\ ,
    \label{Paris3}
  \end{equation}
and this in turn is equivalent to the existence of a set of orthogonal projections $R_\alpha(t_f)$ for any $t_f\geq t_n$
that extend consistently the given set of histories and are perfectly correlated with the histories of the original set. 
The existence of such `generalised' records (which need not be stored in separate degrees of freedom, such as an environment 
or measuring apparatus) is thus equivalent in the case of pure states to strong decoherence (Gell-Mann and Hartle 1990). 
Similar results involving imperfectly correlated records can be derived in the case of mixed states (Halliwell 1999).
The notion of permanent records is rather close to the notion of a `fixed 
past', but the weak decoherence condition will mostly suffice for our purpose of discussing whether probabilities in 
decoherent histories have some genuine time-directed aspect.

\section{Time (a)symmetry and decoherent histories}\label{asymmetry}
\subsection{Standard analysis}
The time asymmetry of the probabilities defined via decoherence consists in the fact that if we exchange the time 
ordering of the histories and insert into (\ref{9}), the resulting histories generally fail to decohere, so one cannot take
the corresponding expression,
  \begin{equation}
    \mbox{Tr}\Big( P_{\alpha_1}(t_1)\ldots 
    P_{\alpha_n}(t_n)\rho P_{\alpha_n}(t_n)\ldots
    P_{\alpha_1}(t_1)\Big)\ ,
    \label{11}
  \end{equation}
to define distribution functions. As we shall see below (see (\ref{Paris12}) and (\ref{Paris13})), the candidate 
probabilities (\ref{11}) are also generally different from those defined by (\ref{8}).

This means that if one wishes to make {\em retrodictions} rather than predictions, i.e.\ 
if one is currently at time $t_{n+1}$ and wishes to calculate the probabilities for the given histories, then one should 
stick to formula (\ref{8}), and not use (\ref{11}). In Schr\"{o}dinger picture, this means 
calculating back the state at $t_0$ and using the predictive formula for that time.  
Indeed, unlike the case of (\ref{8}), where in the Schr\"{o}dinger picture the state 
enters the probability formula as an 
`initial' state (\ref{8A}), in the case of (\ref{11}) the state enters as a `final' state,
  \begin{equation}
    \mbox{Tr}\Big( P_{\alpha_1}U^*_{t_2t_1}\ldots 
    P_{\alpha_n}U^*_{t_{n+1}t_n}\rho(t_{n+1})               
    U_{t_{n+1}t_n} P_{\alpha_n}\ldots U_{t_2t_1}
    P_{\alpha_1}\Big)\ .
    \label{11A}
  \end{equation}
Phenomenologically the quantum state collapses, and if we wish to make retrodictions, we do not use the 
uncollapsed state at time $t_{n+1}$ but the collapsed one (and we often post-select, i.e.\ conditionalise on 
$P_{\alpha_n}$). 

Such time asymmetry in the probabilities is what we actually observe and aim to describe, but in the standard literature 
on decoherent histories (e.g.\ Gell-Mann and Hartle 1994, Kiefer 1996, Hartle 1998) one objects to the fact 
that the asymmetry appears to be inherent in the form of (\ref{9}), i.e.\ decoherent histories simply appear to  
incorporate the asymmetry of the collapse postulate into the fundamental concepts of the approach. We shall now see 
how this problem is usually approached, then develop our own alternative approach.

What is generally done is to {\em modify} the decoherence 
condition in order to obtain a time-neutral criterion. The new condition is that
  \begin{equation}
    \mbox{Re}\mbox{Tr}\Big(\rho_f P_{\alpha_n}(t_n)\ldots 
    P_{\alpha_1}(t_1)\rho_i P_{\alpha'_1}(t_1)\ldots
    P_{\alpha'_n}(t_n)\Big)=0
    \label{14A}
  \end{equation}
for $\{\alpha_i\}\neq\{\alpha'_i\}$, i.e.\ that a new functional should vanish for any two different histories. 
(A condition of this form, and not one of the form (\ref{9}), is actually the one used by Griffiths 1984, who mentions
its explicit time symmetry.)
Note that in this formula there appear {\em two} quantum states $\rho_i$ and 
$\rho_f$, one in `initial' position and one in `final' position. Correspondingly, the probability for a history from a 
set that is decoherent according to the new definition, should be proportional to 
  \begin{equation}
    \mbox{Tr}\Big(\rho_f P_{\alpha_n}(t_n)\ldots 
    P_{\alpha_1}(t_1)\rho_i P_{\alpha_1}(t_1)\ldots
    P_{\alpha_n}(t_n)\Big)\ . 
    \label{14}
  \end{equation}
The normalisation factor can be shown to be equal to $\mbox{Tr}(\rho_f\rho_i)$ (that is, provided this is non-zero).
Indeed, one has
  \begin{multline}
      \mbox{Tr}\Big(\rho_f P_{\alpha_n}(t_n)\ldots P_{\alpha_1}(t_1)\rho_i P_{\alpha_1}(t_1)\ldots P_{\alpha_n}(t_n)\Big) \\
    \begin{split}
      = &
      \mbox{Re}\mbox{Tr}\Big(\rho_f P_{\alpha_n}(t_n)\ldots P_{\alpha_1}(t_1)\rho_i P_{\alpha_1}(t_1)\ldots P_{\alpha_n}(t_n)\Big)
      \\
      = &
      \mbox{Re}\mbox{Tr}\Big(\rho_f P_{\alpha_n}(t_n)\ldots P_{\alpha_1}(t_1)\rho_i 
      (\sum_{\alpha'_1}P_{\alpha'_1}(t_1))\ldots (\sum_{\alpha'_n}P_{\alpha'_n}(t_n))\Big)\ ,
    \end{split} 
    \label{Paris4}
  \end{multline}
because of (\ref{14A}), so that
  \begin{multline}
      \sum_{\alpha_1,\ldots ,\alpha_n}\mbox{Tr}\Big(\rho_f P_{\alpha_n}(t_n)\ldots P_{\alpha_1}(t_1)\rho_i P_{\alpha_1}(t_1)\ldots P_{\alpha_n}(t_n)\Big) \\
    \begin{split}
      = &
      \mbox{Re}\mbox{Tr}\Big(\rho_f (\sum_{\alpha_n}P_{\alpha_n}(t_n))\ldots (\sum_{\alpha_1}P_{\alpha_1}(t_1))\rho_i \Big)
      \\
      = &
      \mbox{Re}\mbox{Tr}(\rho_f \rho_i )\ ,
    \end{split} 
    \label{Paris5}
  \end{multline}
and $\mbox{Tr}(\rho_f \rho_i )$ is real.

By the cyclicity of the trace, it is manifest that 
(\ref{14A}) and the resulting probabilities are indeed time-symmetric, e.g. 
  \begin{multline}
    \frac{\mbox{Tr}\Big(\rho_f P_{\alpha_n}(t_n)\ldots 
    P_{\alpha_1}(t_1)\rho_i P_{\alpha_1}(t_1)\ldots
    P_{\alpha_n}(t_n)\Big)}{\mbox{Tr}(\rho_f \rho_i )}=  \\[1ex]
    \frac{\mbox{Tr}\Big(\rho_i P_{\alpha_1}(t_1)\ldots
    P_{\alpha_n}(t_n)\rho_f P_{\alpha_n}(t_n)\ldots 
    P_{\alpha_1}(t_1)\Big)}{\mbox{Tr}(\rho_i \rho_f )}\ . 
    \label{Paris5B}
  \end{multline}

Thus, according to the standard line, it is necessary to introduce the pair of states 
$\rho_i$ and $\rho_f$ in order to make decoherence not intrinsically time-directed. The observed 
time-asymmetric phenomena are to be explained accordingly in terms of a contingent asymmetry between the 
`initial' and `final' boundary conditions, with $\rho_i$ a certain kind of pure state and $\rho_f$ close to the identity 
operator. The problem is thus reduced to a form familiar from other branches of physics, namely one tries to reduce 
time-directed phenomena to the existence of special boundary conditions (which may or may not be in need of further 
explanation, according to one's take on the problem).

Formula (\ref{Paris5B}) is obviously reminiscent of the ABL formula (\ref{new10}) of standard quantum mechanics (with the 
simplified normalisation factor deriving from the decoherence condition (\ref{14A})), and appears to have been inspired 
by it. Since the decoherent histories formalism is generally considered 
a no-collapse formalism, the two states $\rho_i$ and $\rho_f$ (when translated into Schr\"{o}dinger picture) 
are presumably a pair of non-collapsing wave functions of the universe that take the place of the single wave 
function of the other no-collapse approaches. 

While this proposal 
appears to solve the problem of the time-asymmetry of the decoherence condition, it should not go unquestioned. 

In the first place, it 
is a major modification of quantum mechanics that at present is not required by empirical considerations (the condition 
$\rho_f\approx\iden$ means that the probabilities are indistinguishable in practice from the `time-directed' ones). 
True, there may be reasons for adopting such a modification of standard quantum mechanics. For instance, 
Gell-Mann and Hartle (1994) use their two-state formulation in order to study the possibility of 
`time-symmetric cosmologies' (we shall return to this point in section~\ref{cosmologies}). Further, Hartle
(1998) suggests that the formula may be useful for encoding some violations of unitarity, or that it may become 
necessary if empirical data start violating the usual quantum mechanical probability formula (i.e.\ if $\rho_f$ is 
{\em not} close to the identity). If, however, one's primary concern is with the time asymmetry of the decoherence 
condition, one should ask oneself whether such a radical solution is necessary.

Indeed, this proposal seems to throw out the baby with the bath water, in the following sense. In the context
of the arrow of time, the motivation for adopting a no-collapse approach is the hope that, by insisting on the 
time-symmetric Schr\"{o}dinger equation as fundamental, one might be able to explain the phenomenological time-asymmetry 
of collapse as an effective description. By requiring two states, however, one renounces this strategy without even putting 
up a fight. This is strange, because the interest of the decoherent histories approach (from the point of view of the
arrow of time) would seem to be precisely that the time-asymmetric probabilities (\ref{new7}) appear in the formalism 
without invoking the time-asymmetric evolution (\ref{new6}) of the state. Indeed, we can consider the evolution of the 
state as given always by the Schr\"{o}dinger equation, with the probability formula (\ref{8A}) encoding how the state 
at different times gives the probabilities for suitable histories defined by the corresponding Schr\"{o}dinger projections 
at the given times. Intuitively, one would thus hope that this formalism describes the emergence of time-directed probabilities 
at the level of histories from the fundamental time-symmetric level of the Schr\"{o}dinger equation.

\subsection{New proposal}\label{newsuggestion}
We now claim against standard wisdom that it is {\em not} necessary to go to a more general theory and introduce 
two quantum states $\rho_i$ and $\rho_f$ in order to restore the time symmetry of the decoherent histories 
framework. We shall argue that the usual theory is already symmetric enough and that any phenomenological asymmetry 
can be encoded in a single state $\rho$.

The apparent difficulty with this is 
that while it seems that,
 by adopting the decoherent histories formalism, we have embedded the asymmetric collapse
probabilities (to be suitably interpreted) into the no-collapse formalism of quantum mechanics, as the `no-collapse
strategy' set out to do,
 this has 
been achieved by imposing a condition on the histories that is itself time-asymmetric. So it appears 
that we may have put in the asymmetry by hand at the stage of defining our criterion for decoherence, and that the
resulting asymmetry is an artefact of the formalism. Indeed, the decoherence condition appears to select 
arbitrarily a direction of time, since the 
expression for the interference terms in  (\ref{9}), i.e.
  \begin{equation}
    \mbox{Re}\mbox{Tr}\Big( P_{\alpha_n}(t_n)\ldots 
    P_{\alpha_1}(t_1)\rho P_{\alpha'_1}(t_1)\ldots
    P_{\alpha'_n}(t_n)\Big)\ ,
    \label{again9}
  \end{equation}
is time-directed: these are the interference terms for the evolution of the wave function {\em towards the future}.
The problem can be phrased as the question: why not require that the interference terms {\em towards the past}, i.e.\ 
if we insert $\rho$ into the formula as a `final state',
  \begin{equation}
    \mbox{Re}\mbox{Tr}\Big( P_{\alpha_1}(t_1)\ldots 
    P_{\alpha_n}(t_n)\rho P_{\alpha'_n}(t_n)\ldots
    P_{\alpha'_1}(t_1)\Big)\ ,
    \label{12}
  \end{equation}
vanish for different histories? The arbitrariness lies in our requiring one condition rather than the 
other. 

If this is the problem, however, the most natural move seems to be {\em not} to choose arbitrarily the 
`forwards'  decoherence condition, i.e.\ (\ref{9}), but entertain the possibility that (\ref{12}) might also 
vanish --- a `backwards decoherence' condition,
  \begin{equation}
    \mbox{Re}\mbox{Tr}\Big( P_{\alpha_1}(t_1)\ldots 
    P_{\alpha_n}(t_n)\rho P_{\alpha'_n}(t_n)\ldots
    P_{\alpha'_1}(t_1)\Big)=0
    \label{Parisnew}
  \end{equation}
for any two different histories --- either concurrently with forwards decoherence or as an alternative to it.
Note that the satisfaction of {\em either} condition, given a state $\rho$ and a set of histories is an objective 
feature of the given state and set of histories. And either condition allows us to define probabilities for sets of 
histories.

If only {\em one} condition is satisfied, either forwards decoherence (\ref{9}) or
backwards decoherence (\ref{Parisnew}), the choice is substantial, but it is not arbitrary: again, it is 
the state and the set of histories themselves that select one direction of time over another. (Note also that the terminology of `forwards' and `backwards' decoherence is purely conventional, and that the two terms could be interchanged.)

Before seeing a concrete example, note that in the case in which {\em both} decoherence conditions are satisfied, one can now show that the probabilities 
(\ref{8}) and (\ref{11}), defined by the two conditions, coincide. Indeed, we have
  \begin{multline}
      \mbox{Tr}\Big( P_{\alpha_n}(t_n)\ldots 
      P_{\alpha_1}(t_1)\rho P_{\alpha_1}(t_1)\ldots
      P_{\alpha_n}(t_n)\Big)=                            \\
      \mbox{Re}\mbox{Tr}\Big( P_{\alpha_n}(t_n)\ldots 
      P_{\alpha_1}(t_1)\rho\Big)\ ,
    \label{13}
  \end{multline}
since the left-hand side is equal to its real part and since
  \begin{multline}
      \mbox{Re}\mbox{Tr}\Big( P_{\alpha_n}(t_n)\ldots 
      P_{\alpha_1}(t_1)\rho P_{\alpha_1}(t_1)\ldots
      P_{\alpha_n}(t_n)\Big)=                            \\
      \mbox{Re}\mbox{Tr}\Big( P_{\alpha_n}(t_n)\ldots 
      P_{\alpha_1}(t_1)\rho(\sum_{\alpha'_1}P_{\alpha'_1}(t_1))\ldots
      (\sum_{\alpha'_n}P_{\alpha'_n}(t_n))\Big),
    \label{my23A}
  \end{multline}
by the (forwards) decoherence condition. Similarly with the time order of projections reversed:
  \begin{multline}
      \mbox{Tr}\Big( P_{\alpha_1}(t_1)\ldots 
      P_{\alpha_n}(t_n)\rho P_{\alpha_n}(t_n)\ldots
      P_{\alpha_1}(t_1)\Big)=                            \\
      \mbox{Re}\mbox{Tr}\Big(\rho P_{\alpha_n}(t_n)\ldots 
      P_{\alpha_1}(t_1)\Big).
    \label{my23B}
  \end{multline}
But the two right-hand sides of (\ref{13}) and (\ref{my23B}) are equal
by the cyclicity of the trace. Therefore, the two probabilities coincide:
  \begin{multline}
    \mbox{Tr}\Big(P_{\alpha_n}(t_n)\ldots 
    P_{\alpha_1}(t_1)\rho P_{\alpha_1}(t_1)\ldots
    P_{\alpha_n}(t_n)\Big)=  \\
    \mbox{Tr}\Big(P_{\alpha_1}(t_1)\ldots
    P_{\alpha_n}(t_n)\rho P_{\alpha_n}(t_n)\ldots 
    P_{\alpha_1}(t_1)\Big)\ . 
    \label{My25B}
  \end{multline}

Now, as an example, let us take again the spin-1/2 particle of section~\ref{arrow}. The particle has 
the initial state $\ket{\psi(t_0)}=\alpha\ket{+_x}+\beta\ket{-_x}$ and is subjected
consecutively to a spin-$x$ measurement at $t_1$ and a spin-$z$ measurement at $t_2$. The final state
of the particle and the two apparatuses at some later time $t_3$ is
  \begin{multline}
    \ket{\psi(t_3)}=\frac{\alpha}{\sqrt{2}}\ket{+_z}\otimes\ket{M^1_+}\otimes\ket{M^2_+}+
                    \frac{\alpha}{\sqrt{2}}\ket{-_z}\otimes\ket{M^1_+}\otimes\ket{M^2_-}+  \\
                    \frac{\beta}{\sqrt{2}}\ket{+_z}\otimes\ket{M^1_-}\otimes\ket{M^2_+}+
                    \frac{\beta}{\sqrt{2}}\ket{-_z}\otimes\ket{M^1_-}\otimes\ket{M^2_-}\ .
    \label{none}
  \end{multline}   
Since the four components of the final state are orthogonal, by virtue of the measurement
records, the histories formed by the projections $P^x_\pm(t_1),P^z_\pm(t_2)$, or, more precisely, by the projections
  \begin{equation}
    P^x_\pm(t_1)\otimes\iden\otimes\iden ,\qquad P^z_\pm(t_2)\otimes\iden\otimes\iden\ ,
    \label{alsonone}
  \end{equation}
form a (forwards) decoherent set. One expects this set of histories to fail to decohere backwards, because the initial state lacks records of the later measurements. To show this, given the definition of (\ref{My25B}) above,
we only need to check that the forwards and backwards candidate probabilities (\ref{8}) and (\ref{11}) are
different. And, indeed, this is generally the case. The relevant probabilities can be calculated from the
reduced states of the particle, that is (in Schr\"{o}dinger picture)
  \begin{equation}
    \begin{array}{lcl}
      \ket{\psi(t_0)} & = & \alpha\ket{+_x}+\beta\ket{-_x}\ ,   \\[2ex]
      \rho(t_3)       & = & {\displaystyle \frac{|\alpha|^2+|\beta|^2}{2}\ket{+_z}\bra{+_z} +
                            \frac{|\alpha|^2+|\beta|^2}{2}\ket{-_z}\bra{-_z} = \frac{1}{2}\iden \ .}
    \end{array}
    \label{Paris11}
  \end{equation}
The {\em forwards} probabilities (\ref{8}) are therefore given by
  \begin{equation}
    \begin{array}{lcl}
      p(P^x_+(t_1),P^z_\pm(t_2)) & = & {\displaystyle |\braket{\psi(t_0)}{+_x}|^2\,|\braket{+_x}{\pm_z}|^2 = \frac{|\alpha|^2}{2}\ , }  \\[2ex]
      p(P^x_-(t_1),P^z_\pm(t_2)) & = & {\displaystyle |\braket{\psi(t_0)}{-_x}|^2|\,\braket{-_x}{\pm_z}|^2 = \frac{|\beta|^2}{2}\ , }
    \end{array}
    \label{Paris12}
  \end{equation}
while the {\em backwards} probabilities are given by
  \begin{equation}
    p(P^z_\pm(t_2),P^x_\pm(t_1)) = \mbox{Tr}\Big(\frac{1}{2}\iden\ket{\pm_z}\bra{\pm_z}\Big)|\braket{\pm_z}{\pm_x}|^2=\frac{1}{4}\ .  
    \label{Paris13}
  \end{equation}
Thus, unless $|\alpha|^2=|\beta|^2=\frac{1}{2}$, the forwards and backwards probabilities do not coincide, and therefore 
the histories fail to decohere backwards in time.

If the {\em final} state had the form $\ket{\psi(t_3)}=\alpha\ket{+_x}+\beta\ket{-_x}$, the reverse would be true, and
for $|\alpha|^2\neq |\beta|^2$ the same set of histories would decohere backwards but not forwards in time. In either case,
satisfaction of only one of the decoherence conditions gives rise to probabilities that are time-asymmetric both in their 
formal expression and in their actual values.

Clearly, to explain this `arrow of decoherence' in general, one should analyse the (symmetric) role played by the dynamics 
(what kind of Hamiltonians are necessary in order to obtain decoherence?) and the (symmetry-breaking) role played by 
special, presumably (as here) `low-entanglement' initial or final conditions (note that initial and final quantum states 
cannot be specified independently since unitary evolution is assumed throughout). Further questions are whether and 
how these conditions relate to the existence of observers and agents like us. But
the discussion in this paper is limited to the formal aspects of the time symmetry and asymmetry of probabilities.
Insofar as the standard proposal in the decoherent histories literature (which we have reviewed in the previous subsection) 
allows one to reduce the time asymmetry of decoherence to a more familiar one of explaining special initial or final 
conditions, so does the present suggestion: depending on an appropriate initial or final condition (corresponding to 
a {\em single} Heisenberg-picture $\rho$), there are certain processes, namely the 
suppression of interference terms between certain histories, that typically occur in one direction of time (\ref{9}) 
or the opposite one (\ref{Parisnew}), although the fundamental equation allows for both types of solutions. Unlike the 
standard approach, our suggestion applies to the framework of the usual Schr\"{o}dinger equation, thus along 
the established lines of the no-collapse approach to the arrow of collapse in quantum mechanics (cf.\ section~\ref{collnocoll}).

\subsection{Time-symmetric case}\label{bothconditions}
We shall now have a brief look at the symmetrical situation in which both the forwards and the
backwards decoherence condition are satisfied. In this case, the 
arbitrariness of choosing one condition over another does not matter. 

We shall show in particular: (i) that the time-symmetric case is non-trivial, in the sense that probabilities for 
histories that decohere in both directions of time need not be all 0 or 1 (even in the case of a pure quantum state), and
(ii) that satisfaction of both forwards and backwards decoherence is not equivalent to the special case of the standard
time-neutral decoherence condition (\ref{14A}) with $\rho_i=\rho_f=\rho$, i.e. it is not equivalent to
  \begin{equation}
    \mbox{Re}\mbox{Tr}\Big(\rho P_{\alpha_n}(t_n)\ldots P_{\alpha_1}(t_1)
                           \rho P_{\alpha'_1}(t_1)\ldots P_{\alpha'_n}(t_n)\Big)=0\
    \label{ParisA}
  \end{equation}
(for different histories).  In this connection, it should be noted that Gell-Mann and Hartle (1994) show (\ref{ParisA}) to be an overly restrictive condition, as well as showing that the assumption that both $\rho_i$ and $\rho_f$ are pure is also very restrictive. 

To show both the above claims,
take the special case of (\ref{ParisA}) with $\rho=\ket{\psi}\bra{\psi}$ pure. In this case, the corresponding
probabilities (\ref{Paris5B}) reduce to
  \begin{multline}
    \mbox{Tr}\Big(\ket{\psi}\bra{\psi} P_{\alpha_n}(t_n)\ldots P_{\alpha_1}(t_1)
                  \ket{\psi}\bra{\psi} P_{\alpha_1}(t_1)\ldots P_{\alpha_n}(t_n)\Big)=  \\
    |\bra{\psi} P_{\alpha_n}(t_n)\ldots P_{\alpha_1}(t_1)\ket{\psi}|^2\ .                       
    \label{nolabelagain}
  \end{multline}
By the same argument used to derive (\ref{Paris4}) or (\ref{my23A}), we see that the same probabilities
are also equal to
  \begin{multline}
    \mbox{Tr}\Big(\ket{\psi}\bra{\psi} P_{\alpha_n}(t_n)\ldots P_{\alpha_1}(t_1)
                  \ket{\psi}\bra{\psi} P_{\alpha_1}(t_1)\ldots P_{\alpha_n}(t_n)\Big)  \\
    \begin{split}
      = & \mbox{Tr}\Big(\ket{\psi}\bra{\psi} P_{\alpha_n}(t_n)\ldots P_{\alpha_1}(t_1)
                  \ket{\psi}\bra{\psi}\Big)  \\
      = & \bra{\psi} P_{\alpha_n}(t_n)\ldots P_{\alpha_1}(t_1)\ket{\psi}\ .
    \end{split}
    \label{andagain}
  \end{multline}
But then, all probabilities are 0 or 1. 

In this case, since the
$U_{t_{n+1}t_n}P_{\alpha_n}U_{t_nt_{n-1}}\ldots P_{\alpha_1}U_{t_1t_0}\ket{\psi(t_0)}$ are orthogonal components of
$\ket{\psi(t_0)}$ (in Schr\"{o}dinger picture), each of the former must be the zero vector in order to be orthogonal 
to the latter. It follows that the
forwards and backwards decoherence conditions are trivially satisfied.
Instead, we shall now show that 
the forwards and backwards decoherence conditions can be satisfied for a pure state also
when the probabilities are not all 0 or 1. 

Take again the special case above of the spin-1/2 particle. In this case, the additional assumption of backwards 
decoherence not only implies (\ref{My25B}), but is also implied by the equality of the 
(candidate) probabilities (\ref{8}) and (\ref{11}). Indeed, assume that
  \begin{multline}
    \mbox{Tr}\Big(P^z_+(t_2)P^x_\pm(t_1)\ket{\psi}\bra{\psi}P^x_\pm(t_1)P^z_+(t_2)\Big)= \\
    \mbox{Tr}\Big(P^x_\pm(t_1)P^z_+(t_2)\ket{\psi}\bra{\psi}P^z_+(t_2)P^x_\pm(t_1)\Big)
    \label{reducendum}
  \end{multline}
(in Heisenberg picture). Then, since (again due to forwards decoherence) the first line is equal to
  \begin{equation} 
    \mbox{Tr}\Big(P^z_+(t_2)P^x_\pm(t_1)\ket{\psi}\bra{\psi}\Big)=
    \mbox{Tr}\Big((P^z_+(t_2)+P^z_-(t_2))\ket{\psi}\bra{\psi}P^z_+(t_2)P^x_\pm(t_1)\Big)\ ,
  \end{equation}
equation (\ref{reducendum})
reduces to 
  \begin{multline}
    \mbox{Tr}\Big((P^z_+(t_2)+P^z_-(t_2))\ket{\psi}\bra{\psi}P^z_+(t_2)P^x_\pm(t_1)\Big)=  \\
    \mbox{Tr}\Big(P^z_+(t_2)\ket{\psi}\bra{\psi}P^z_+(t_2)P^x_\pm(t_1)\Big)\ .
    \label{reductum}
  \end{multline}
Therefore
  \begin{equation}
    \mbox{Tr}\Big(P^x_\pm(t_1)P^z_-(t_2)\ket{\psi}\bra{\psi}P^z_+(t_2)P^x_\pm(t_1)\Big)=0
  \end{equation}
and {\em a fortiori}
  \begin{equation}
    \mbox{Re}\mbox{Tr}\Big(P^x_\pm(t_1)P^z_-(t_2)\ket{\psi}\bra{\psi}P^z_+(t_2)P^x_\pm(t_1)\Big)=0\ ,
  \end{equation}
which is backwards decoherence.

Since in the case $|\alpha|^2=|\beta|^2=\frac{1}{2}$ the probabilities (\ref{Paris12}) and (\ref{Paris13})
coincide, we have an example in which both forwards and backwards decoherence are satisfied for a pure state, 
showing both that probabilities can be non-trivial and that (\ref{ParisA}) is not equivalent to the conjunction of 
forwards and backwards decoherence.

Note that the example just cited hardly qualifies as a physically interesting 
example of satisfaction of both conditions. Indeed, while forwards (strong) decoherence in the example is connected
to the existence of measurement records, backwards (strong) decoherence in the example appears to be an accident,
and the corresponding records are indeed merely `generalised' records. (Note also that whether this qualifies as a 
case of intuitive `branching' of the wave function in both directions of time may be open to doubt.)

\subsection{Time-symmetric cosmologies}\label{cosmologies}
To conclude this section, we wish to note that our time-symmetric case (satisfaction of both forwards and backwards
decoherence) is also in general not equivalent to what Gell-Mann and Hartle (1994) take as the defining characteristic of
a `time-symmetric cosmology', namely decoherence and equality of probabilities for both a set of histories
  \begin{equation}
    P_{\alpha_1}(t_1),\ldots,P_{\alpha_n}(t_n)
    \label{star}
  \end{equation}
and the {\em time-reversed} set of histories
  \begin{equation}
    P_{\alpha_n}(-t_n),\ldots,P_{\alpha_1}(-t_1)
    \label{doublestar}
  \end{equation}
(more precisely, Gell-Mann and Hartle consider CPT-reversed sets of histories). 

Taking our case of a single $\rho$ (or $\rho_i=\rho$ and $\rho_f=\iden$ in Gell-Mann
and Hartle's formulas), we have however that if $\rho$ is time-symmetric, then (quite trivially) forwards decoherence
of the time-reversed set of histories (\ref{doublestar}), is equivalent to backwards decoherence of the original set 
(\ref{star}). From our discussion in section~\ref{newsuggestion} it then also follows that the probabilities for 
(\ref{star}) and (\ref{doublestar}) coincide, i.e.\ time symmetry of $\rho$ and our two decoherence conditions 
imply a time-symmetric cosmology in Gell-Mann and Hartle's sense.

More generally, Page (1993) has shown that if $\rho_i$ and $\rho_f$ in Gell-Mann and Hartle's formulas are separately
time-symmetric and commute, then decoherence of both (\ref{star}) and (\ref{doublestar}) in the sense of (\ref{14A})
implies that their probabilities coincide and thus that one has a time-symmetric cosmology. The proof is quite similar to our proof of (\ref{My25B}), and in fact if $\rho_i$ and
$\rho_f$ are time-symmetric, then the `time-neutral' decoherence functional applied to (\ref{doublestar}), 
  \begin{equation}
    \mbox{Re}\mbox{Tr}\Big(\rho_iP_{\alpha_n}(-t_n)\ldots P_{\alpha_1}(-t_1)\rho_fP_{\alpha'_1}(-t_1)\ldots P_{\alpha'_n}(-t_n)\Big)\ , 
  \end{equation}
is equal to
  \begin{equation}
    \mbox{Re}\mbox{Tr}\Big(\rho_iP_{\alpha'_n}(t_n)\ldots P_{\alpha'_1}(t_1)\rho_fP_{\alpha_1}(t_1)\ldots P_{\alpha_n}(t_n)\Big)\ , 
  \end{equation}
which is so to speak a two-state version of the backwards decoherence functional applied to the original histories. 

Page (1993) stresses that time-symmetric cosmologies in the sense of Gell-Mann and Hartle can thus be obtained even without requiring a sufficient condition mentioned by these authors, namely that $\rho_i$ and $\rho_f$ be the time reversals of each other (which given the rest of their discussion would most likely rule out pure states, such as in a no-boundary cosmology). In particular one can obtain time-symmetric cosmologies even if one takes $\rho_f=\iden$. We should like to add that, therefore, the wish to consider time-symmetric cosmologies does also not provide a reason for introducing Gell-Mann and Hartle's two-state condition (\ref{14A}).

Note finally that any example of a time-symmetric cosmology in the sense of Page with $\rho_i$ pure and time-symmetric
and $\rho_f$ the identity will automatically provide an example of our time-symmetric case of a set of histories satisfying 
both forwards and backwards decoherence.

\section{Discussion of probabilities}\label{discussion}
We now return to our original question of assessing the possibly time-directed metaphysical status
of quantum probabilities  in the context of (decoherence-based) no-collapse approaches to quantum mechanics. 

So far we have conducted a formal discussion of the time symmetry or asymmetry of
these probabilities in terms of forwards and backwards decoherence. 
We have argued in section~\ref{asymmetry} that the framework of decoherent histories, which we use here to describe no-collapse approaches, provides a way of embedding the quantum probabilities within the usual time-symmetric description given by the Schr\"{o}dinger equation, in a way that does not beg the question of time directedness. Probabilities emerge from the Schr\"{o}dinger equation at the level of histories, contingently on the satisfaction of some decoherence condition. Sometimes the emergent probabilities are time-asymmetric, in the sense that the corresponding histories satisfy only one of the two decoherence conditions we have introduced. Sometimes they are time-symmetric, in the sense that the histories satisfy both decoherence conditions, in which case they appear to be no more time-directed than classical probabilities.

As already mentioned in section~\ref{probs}, while formal considerations cannot enforce the choice of a particular kind of interpretation, they may very well provide guidance in the choice of an appropriate interpretation.   Our formal results suggest caution in making sweeping metaphysical statements about the nature of quantum probabilities, in particular about fixed past and open future, suggesting instead that any time-directed aspect one may want to ascribe to specifically quantum probabilities when fleshing out  their interpretation (in the context of some no-collapse approach to quantum mechanics, or if invoking the theorem about records to justify some kind of fixed past) should be thought of as merely contingent.

A second type of considerations revolves not around forwards and backwards decoherence but around decoherence and recoherence, which is the {\em reinterference} at later times of components of the quantum state that satisfied the decoherence condition in the past.  

For example, choose $t_1<\ldots<t_n<0$. 
While a set of histories may satisfy the forwards decoherence condition,
  \begin{equation}
    \mbox{Re}\mbox{Tr}\Big( P_{\alpha_n}U_{t_nt_{n-1}}\ldots 
    P_{\alpha_1}U_{t_1t_0}\rho(t_0)                 
    U^*_{t_1-t_0} P_{\alpha'_1}\ldots U^*_{t_nt_{n-1}}
    P_{\alpha'_n}\Big)=0
  \end{equation}
(for different histories), this does not prevent reinterference from taking place at times later than $t_n$.
Indeed, take the extreme case of a time-symmetric $\rho$ (in the pure case, this corresponds to (\ref{4}) or (\ref{5})). Then the corresponding components of the wave function will progressively reinterfere at the times $0<-t_n<\ldots<-t_1$.  
If a strongly decoherent set of histories recoheres, then the records at times $t_i$ of all previous events will not be permanent, but will be subject to successive `quantum erasure' between $-t_n$ and $-t_1$.

Note that recoherence in the example is equivalent to the backwards decoherence of the time-reversed set of histories $P_{\alpha_n}(-t_n),\ldots,P_{\alpha_1}(-t_1)$, which is different from backwards decoherence of the original set of histories. Therefore, issues of forwards and backwards decoherence are indeed separate from issues of decoherence and recoherence.

The possibility of recoherence now raises an important potential objection to the time-directedness of quantum probabilities. This objection is already known from classical discussions of the thermodynamic arrow of time: any apparent emergence of an arrow of time (say through the imposition of an initial low-entropy boundary condition) can be overturned by behaviour in the future (e.g.\ it could be trumped by a final low-entropy boundary condition). This problem has been discussed in particular by Price (2002), who refers to it picturesquely as `Boltzmann's time bomb'. Thermodynamic behaviour could turn into anti-thermodynamic behaviour, for instance in a perfectly symmetric universe, as in the quantum case of collapse behaviour turning into `anti-collapse' behaviour in the above example with a time-symmetric $\rho$. Therefore the emergence of probabilities, even if asymmetric in the sense discussed in section~\ref{asymmetry}, is in general a temporally local phenomenon, again suggesting caution in one's choice of interpretation.

There is, however, a feature of the quantum probabilities under recoherence that provides a disanalogy to the classical case. Consider whether anti-thermodynamic behaviour could be actually witnessed by some `thermodynamic' observers (i.e.\ observers associated with the thermodynamic arrow of time). This will not be the case in the extreme example of a perfectly symmetric universe, since also the operation of their observations and memories will be reversed (that is, they will be acting as observers of the thermodynamic behaviour in the opposite time direction). However,  one can imagine that in the presence of sufficiently weak interactions, some thermodynamic observers  could 
directly witness some of the anti-thermodynamic behaviour.

This now yields a disanalogy between the classical and the quantum case. In  the quantum case, even in the more general situation in which the state $\rho$ is not time-symmetric, no observer could arguably ever witness decoherence events followed by recoherence events at some later time, if by observing an event in a history one means that the observer subsequently possesses a record of this event. Indeed, as long as records of events persist in the memory of the observer, the histories comprising the observed events will decohere, in fact strongly, because of the theorem about records. The assumption of recoherence therefore implies that the observer's memory of an event be quantum-erased. If this is the case, however, 
that observer arguably cannot be said to witness a recoherence event: once the different components of the quantum state have come together, the observer (if they survive the process) has no memory of them ever having been distinct, thus cannot be said to witness an event of distinct components of the quantum state reinterfering.

While for an external `God's eye' perspective (such as we arguably have in spin-echo experiments)decoherence and recoherence are equivalent time-reversed descriptions of the same phenomenon , there can be no internal observer who shares the temporal perspective from which this behaviour
could be described as recoherence. From an internal perspective such a phenomenon could only be described as decoherence. In this specifically perspectival sense, the arrow of decoherence is immune to special final conditions in its future. This in turn provides scope for an interpretation in which the quantum probabilities can be thought of as time-directed, not as a property of the probabilities themselves, but from the perspective of any observer in the above sense.

The interpretation of quantum probabilities will in general depend on the particular approach to quantum mechanics one adopts, in particular, even restricting oneself to no-collapse approaches, on whether one 
adopts a (deterministic) Bohm approach, a (stochastic) beable approach or an (emergentist) Everett approach. Our analysis above suggests that in all of these cases, the interpretation of quantum probabilities should take into account both the dichotomy between forwards and backwards decoherence, suggesting that time directedness should be contingent, and the dichotomy between decoherence and recoherence, suggesting that time directedness is inappropriate from a global perspective, but may be construed as genuine from the perspective of an internal observer.

\section*{Acknowledgements}

This paper has a long history, the first ideas having been conceived during the workshop on {\it Real Time and its Quantum Roots} at the University of Arizona at Tucson in April 2001. Preliminary versions of this material were presented at the Oxford Philosophy of Physics Seminar in June 2001, the Philosophy of Physics Saturdays, Berkeley, November 2001, the Seminario Interdisciplinare (Filosofia-Fisica), Dipartimento di Fisica, Universit\`{a} di Firenze, March 2003, and the workshop on {\em Quantum Mechanics on the Large Scale}, University of British Columbia, April 2003. More mature versions were presented at the conference on {\em Probability in Quantum Mechanics}, LSE, February 2004, at the Institut f\"{u}r Grenzgebiete der Psychologie und Psychohygiene (IGPP), Freiburg, May 2004, at the IGPP-BION interdepartmental meeting, Beuggen, also in May 2004, at the Paris Philosophy of Physics Seminar, November 2005, and at the Summer Foundations Workshops at the Centre for Time, University of Sydney, December 2005 and 2006. The first version of this paper was written while I was holding an Alexander von Humboldt Fellowship at IGPP in Freiburg. The final version is heavily indebted to the excellent suggestions by Bernard d'Espagnat, Roman Frigg and Stephan Hartmann, and Huw Price. My thanks go also to David Albert, Hilary Greaves, Max Schlosshauer, Lev Vaidman, two anonymous referees and all others who have contributed to sharpen the content of this paper through comments or discussion.

\section*{References}

Aharonov, Y., Bergmann, P.\ G., and Lebowitz, J.\ L.\ (1964), `Time Symmetry in the Quantum Process of Measurement', 
{\em Physical Review} {\bf 134}, B1410-B1416.

Aharonov, Y., and Vaidman, L.~(2002), `The Two-State Vector Formalism of Quantum Mechanics', in 
J.~G.~Muga, R.~Sala Mayato and I.~L.~Egusquiza (eds.), {\em Time in Quantum Mechanics}, pp.~369--412, quant-ph/0105101~. 

Bacciagaluppi, G.\ (2002), `Remarks on Space-time and Locality in Everett's Interpretation', in T.~Placek and J.~Butterfield 
(eds.), {\em Non-locality and Modality}, NATO Science Series II. Mathematics, Physics and Chemistry, Vol.~64 (Dordrecht: Kluwer),
pp.~105--122, PHILSCI-00000504~.

Bacciagaluppi, G.\ (2003), `The Role of Decoherence in Quantum Mechanics', in E.~N.~Zalta (ed.), {\em The Stanford
Encyclopedia of Philosophy},\\ plato.stanford.edu/archives/win2003/entries/qm-decoherence~.

Bell, J. S.\ (1984), `Beables for Quantum Field Theory', in {\em Speakable and Unspeakable in Quantum 
Mechanics} (Cambridge: CUP, 1987), pp.~173--180.

Dowker, F., and Kent, A.~(1996), `On the Consistent Histories Approach to Quantum Mechanics', 
{\em Journal of Statistical Physics}, {\bf 82}, 1575--1646.

D\"{u}rr, D., Goldstein, S., and Zangh\`{\i}, N.\ (1992), `Quantum Equilibrium
and the Origin of Absolute Uncertainty', {\em Journal of Statistical Physics} {\bf 67}, 843--907.

Gell-Mann, M., and Hartle, J.~B.\ (1990), `Quantum Mechanics in the Light of Quantum Cosmology', in W.~H.~Zurek (ed.),
{\em Complexity, Entropy, and the Physics of Information} (Reading: Addison-Wesley), pp.~425--458.

Gell-Mann, M., and Hartle, J.~B.\ (1994), `Time Symmetry and Asymmetry in Quantum Mechanics and Quantum Cosmology', in
J.~J.~Halliwell, J.~P\'{e}rez-Mercader and W.~H.~Zurek (eds.), {\em Physical Origins of Time Asymmetry} (Cambridge: 
CUP), pp.~311--345. 

Griffiths, R.~B.\ (1984), `Consistent Histories and the Interpretation of Quantum Mechanics', {\em Journal of Statistical
Physics} {\bf 36}, 219--272.


Halliwell, J.~J.\ (1999), `Somewhere in the Universe: Where is the Information Stored when Histories Decohere?',
{\em Physical Review} {\bf D 60}, 105031-1--17, quant-ph/9902008~.

Hartle, J.\ B.\ (1998), `Quantum Pasts and the Utility of History', {\em Physica Scripta} {\bf T76}, 67--77,
gr-qc/9712001~.

Kiefer, C.\ (1996), `Consistent Histories and Decoherence', chapter~5 in D.~Giulini, E.~Joos, C.~Kiefer, J.~Kupsch, 
I.-O.~Stamatescu and H.~D.~Zeh, {\em Decoherence and the Appearance of a Classical World in Quantum Theory} (Berlin: 
Springer), pp.~157--186.

Omn\`{e}s, R.\ (1988), `Logical Reformulation of Quantum Mechanics, I--III', {\em Journal of Statistical
Physics} {\bf 53}, 893--975.

Page, D.\ (1993), `No Time Asymmetry from Quantum Mechanics', {\em Physical Review Letters} {\bf 70}, 4034--4037. 

Price, H.\ (2002), `Boltzmann's Time Bomb', {\em British Journal for the Philosophy of Science} {\bf 53}, 83--119.

Saunders, S.\ (1993), `Decoherence, Relative States, and Evolutionary Adaptation', {\em Foundations of Physics}
{\bf 23}, 1553--1585.

Vaidman, L.~(2007), `Backward Evolving Quantum States', to appear in {\em Journal of Physics A}, quant-ph/0606208~. 

Wallace, D.~(2005), `Quantum Probability from Subjective Likelihood: Improving on Deutsch's proof of the Probability
Rule', to appear in {\em Studies in History and Philosophy of Modern Physics}, PHILSCI-00002302~.

\end{document}